# Wearable Health Monitoring Using Capacitive Voltage-Mode Human Body Communication

Shovan Maity, Student Member, IEEE, Debayan Das, Student Member, IEEE and Shreyas Sen, Member, IEEE
School of Electrical and Computer Engineering, Purdue University
{maity, das60, shreyas}@purdue.edu

*Abstract*— Rapid miniaturization and cost reduction of computing, along with the availability of wearable and implantable physiological sensors have led to the growth of human Body Area Network (BAN) formed by a network of such sensors and computing devices. One promising application of such a network is wearable health monitoring where the collected data from the sensors would be transmitted and analyzed to assess the health of a person. Typically, the devices in a BAN are connected through wireless (WBAN), which suffers from energy inefficiency due to the high-energy consumption of wireless transmission. Human Body Communication (HBC) uses the relatively low loss human body as the communication medium to connect these devices, promising order(s) of magnitude better energy-efficiency and built-in security compared to WBAN. In this paper, we demonstrate a health monitoring device and system built using Commercial-Off-The-Shelf (COTS) sensors and components, that can collect data from physiological sensors and transmit it through a) intra-body HBC to another device (hub) worn on the body or b) upload health data through HBC-based human-machine interaction to an HBC capable machine. The system design constraints and signal transfer characteristics for the implemented HBC-based wearable health monitoring system are measured and analyzed, showing reliable connectivity with >8x power savings compared to Bluetooth low-energy (BTLE).

I. INTRODUCTION

Five decades of continuous scaling following Moore's law has led to constant shrink in size and cost of unit computing. This has enabled cheap, ubiquitous computing to be incorporated in small form-factor devices. On the other hand, rapid advancement in wireless communication technology has made it possible to connect these small computation units seamlessly, forming an internet of billions of devices. The emergence of cheap connected computing has resulted in the modern era of interconnected smart devices also commonly known as the Internet of Things (IoT). In addition, there has been widespread availability of cheap on-body and implantable sensors like Electrocardiography (ECG), Blood pressure, Glucose sensors to monitor vital physiological parameters. With the increase in the world's aging population and the aggregated costs of hospitalization and patient care, there is an urgent need for providing new methods of health monitoring and diagnostic systems at affordable cost. Therefore, the prolific growth of miniaturized sensors coupled with continued miniaturization of unit computing, and the need for affordable healthcare; wearable/implanted physiological sensors are becoming increasingly common in medical, healthcare systems and for remote health monitoring.

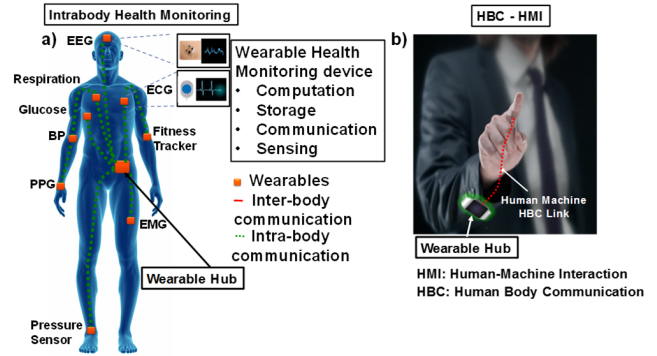

Figure 1: Wearable Health Monitoring using Human Body Communication (HBC) a) Remote Health monitoring network with multiple on body sensors exchange health data with a Wearable Hub through HBC. b) HBC-HMI: Human Machine Interaction through dynamically formed HBC channel during Human Machine Interaction can that can be used to securely upload data from the Wearable Hub to a medical hub/device for further monitoring.

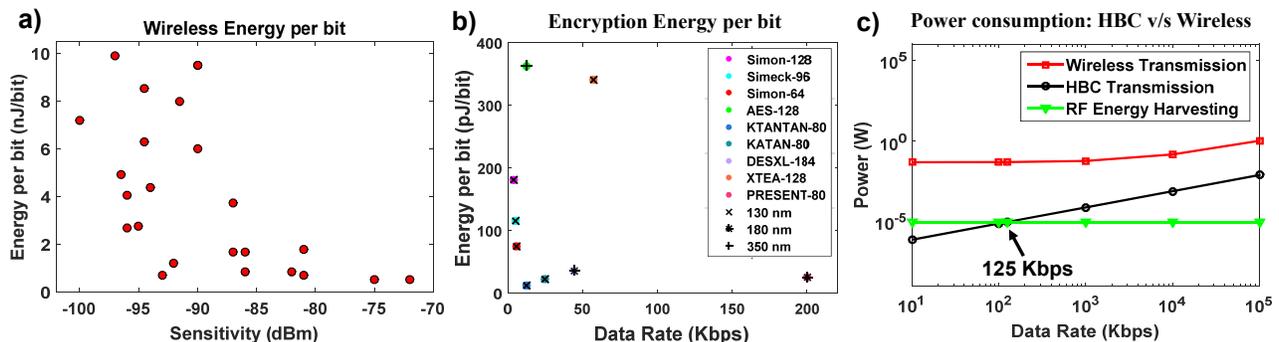

Figure 2: a) Energy consumption of wireless transmission with different receiver sensitivity showing energy consumption in nJ/bit [7]. b) Encryption energy for different lightweight cryptographic implementations [3]-[6]. c) Power consumption comparison between wireless and HBC systems for different data rate taking in values from a), b). It shows HBC power consumption is ~2 orders less compared to wireless transmission enabling energy-harvested nodes.

Figure 1 shows the proposed HBC based remote wearable health monitoring system consisting of physiological sensor nodes and a Wearable Hub (WH). Health data from several wearable/implantable physiological sensors is collected by an on-body hub using HBC (Figure 1a). A metal electrode is used to couple the signal to the human body and a similar electrode receives the signal from the body at the WH. The WH aggregates the data, encrypts it and transmits it through a wireless medium to a remote station. The WH has a bigger battery, hence it is less energy constrained and may have wireless transmission capability with encryption (to keep private health data secure). Another option (Figure 1b) is to store the collected health data on WH and securely upload it to a health-terminal through HBC on a daily basis at home, or work, or whenever the user visits the hospital. The WH can transmit the data through HBC when the user touches an electrode in the receiving machine enabling human-machine interaction through HBC.

Limited energy availability due to small battery size is one of the primary constraints of the devices of a remote health monitoring system. For perpetual operation of wearable/implantable physiological sensors, in near future, they need to be powered through harvested energy (Radio Frequency (RF), thermal, motion, vibration etc.). This requires the computation and communication within the nodes to be performed with extremely low power consumption. Since communication power can be a significant portion of the total power consumption in a computationally inexpensive sensor node it is of utmost importance to have a communication medium, which enables ultra-low power communication. Traditional state-of-the-art method of connecting these physiological sensors is through a Wireless Body Area Network (WBAN). However the energy required for traditional wireless transmission is >1-10 nJ/bit [1]. Figure 2a shows the energy consumption of wireless communication, Figure 2b shows the encryption energy and Figure 2c shows the relative comparison between HBC and wireless energy consumption [2]–[6]. From Figure 2c, it is evident that from RF energy harvesting we cannot have enough energy to support wireless communication in the physiological sensor nodes. However, Human Body Communication (HBC) can provide nearly two orders higher energy efficiency compared to wireless transmission of ~100 pJ/bit [7]. This will allow us to communicate up to a certain data rate even on energy harvested sensor nodes, enabling a HBC based health monitoring system. In wireless transmission (WBAN), the signal is broadcasted in the wireless medium and can be snooped by an attacker from free space making it necessary to encrypt the data before transmission. On the other hand HBC will enable data transfer with lesser chance of snooping from a nearby attacker, since the signals are mostly confined within the human body. Hence adopting HBC as transmission medium will provide higher security compared to WBAN and enable low energy physiological sensor based health-monitoring system.

The key contributions of this paper are as follows:
- **HBC-based Wearable Health Monitoring System (WHMS) with COTS components**: A COTS based implementation of secure, energy-efficient WHMS using HBC is demonstrated with performance measurement and analysis showing >8x benefit over WBAN. The COTS platform for HBC is enabled by a) General Purpose Input Output (GPIO) based wireline transmitter and ADC based wireline receiver along with b) low-loss capacitive HBC.
- **Low-Loss Capacitive Voltage-Mode (VM) HBC**: Use of Capacitive VM HBC allows for lower-loss HBC transmission, increasing SNR. VM HBC refers to communication using a high input impedance receiver. The signal return path is provided by capacitive coupling between the transmitter, receiver and earth's ground. Capacitive VM HBC uses a high impedance capacitive receiver with capacitive return path to allow a wideband lower-loss HBC channel.
- **Human-Machine Interaction through Dynamic HBC (HBC-HMI)**: We demonstrate the framework for health data upload from WH to Health Monitoring Machine using dynamic HBC-HMI.

The rest of the paper is organized as follows. Section II talks about the signal transfer characteristics of HBC, and how this can be used for health monitoring. Section III discusses about a prototype design with COTS components showing Intra-body and Human-Machine data transfer. Section IV looks into some of the important signal characteristics of the system with conclusion in Section V.

## II. HUMAN BODY COMMUNICATION

Intra-body HBC was first proposed as a method of connecting devices on a Personal Area Network in [8] . Both the receiver and transmitter have a pair of electrodes, are electrically isolated and battery powered. The transmitter capacitively couples the signal into the human body creating a small displacement current, which is picked up at the receiver end. This is an example of HBC through *Capacitive coupling*. Wegmueller *et al.* introduced *Galvanic coupling* [9] where the transmitted signal is applied between two electrodes directly connected to the human body thus inducing an alternating electric current with peak amplitude of ~1 mA, propagating through the body. At the receiver end, there are two electrodes, which sense the potential difference created due to the induced electric field created from the current. However, in Galvanic coupling as the two electrodes in the transmitter end forms a relatively low resistance path for electric current to flow, most of the current flows in this path. This increases the loss in galvanic coupling drastically as the distance between the transmitter and receiver increases. Because of high loss, a galvanic coupling based reliable COTS based implementation is extremely hard. However, as COTS platform provides the advantage of fast proof-of-concept demonstration, in this paper we demonstrate a HBC based WHMS system using COTS and analyze its SNR and energy-efficiency benefits over WBAN based WHMS. Using capacitive coupling and a high input impedance voltage-mode receiver, the loss at the receiver end can be reduced significantly enabling COTS based implementation. Although the capacitive voltage-mode HBC loss is lower compared to galvanic coupling, but the voltage loss is high enough to reduce the received signal amplitude to such a level that it cannot be treated as a digital signal at the receiver end. This requires the signal to be sampled through an ADC as an analog signal at the receiver end, the details for which is discussed in the following section.

Previously, human machine interaction through HBC has been demonstrated using Galvanic Coupling. In this paper, we demonstrate capacitive VM human machine interaction (lower loss, better reliability), for the first time, as a method of uploading data from WH to external device.

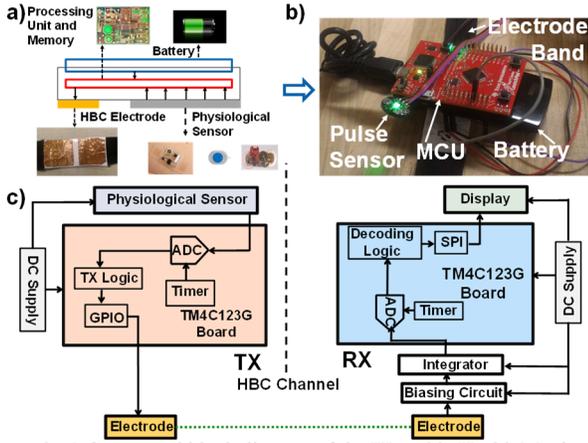

Figure 3: a) Conceptual block diagram of the Wearable Health Monitoring Device. b) Actual Picture of the device showing different components. c) System level diagram showing interaction between different system blocks.

## III. HBC BASED HEALTH MONITORING

### A. System Nodes

A HBC based health monitoring system consists of multiple physiological sensors and a Wearable Hub (WH). The sensor nodes collect physiological sensor data, process it and communicate it through HBC to the WH. The sensor nodes have very limited battery life or are energy harvested. The WH collects data from all sensors in the health monitoring system, aggregates and sends them for further analysis and diagnosis. The data transmission can happen either through wireless transmission or through Human-Machine HBC when the user comes in contact with a machine having HBC receiving capability. The later method will have higher energy efficiency compared to wireless transmission and promises physically secure communication. In this paper, we focus on building the HBC framework that will allow us to enable a) intra-body communication (sensor to WH) and b) human-machine interaction (WH to machine).

### B. Health Monitoring Node Design

A Wearable Health Monitoring device has been built which communicates using the human body as the communication medium. The goal is to build a wearable device, which can interface a sensor to collect data and transmit it through HBC. To this end, the device should take external input, process, store the data and be able to communicate it. The system level, conceptual block diagram and actual implementation is shown in Figure 3.

Using mostly off-the-shelf blocks and a few custom designed blocks, the Health Monitoring device has been designed. The conceptual diagram (Figure 3a) shows that the device consists of a communication module, processing module, memory, power source, sensor and an interface with the human body. The communication module is used to send and receive data. The processing module processes external inputs coming from sensors, performs operations on the received data. The device is battery powered to emulate the scenario of a wearable device. An interface electrode is required to couple the transmitted data into the body and also receive the data transmitted through the body. A Texas Instruments *TM4C123G* LaunchPad evaluation kit consisting of an ARM Cortex M4 based *TM4C123GH6PM* microcontroller is used for implementing the communication and processing module together. A Duracell rechargeable Lithium ion battery is used as power supply. The coupling of the signals between the microcontroller and the human body is done through bands consisting of copper electrodes. As a physiological sensor we have used a wearable pulse sensor for non-invasive heart rate monitoring.

### C. Wearable Biomedical Monitoring System Design

Since the HBC receiver and the transmitter do not share a common ground only AC signals can be transmitted through the human body. In addition, the signal at the receiver end is attenuated due to weak return path, so standard serial protocols such as UART, SPI, and I2C cannot be used at the receiver end. To solve this problem, the received AC signal is coupled onto a DC level set by a resistive divider at the receiver by capacitive coupling. The received AC signal is attenuated and only has a peak-to-peak amplitude of 45 mV, for a 3.3 V peak-to-peak transmitted signal. We integrate and sample the received signal using a 12 bit 1MSPS ADC present in the controller and the logic level is determined by comparing the digitized value of the received signal with the previous sample value, and looking at their difference. Figure 4 describes the characteristics of a few important signals at the system level.

The frequency of operation of the system is constrained by several factors. The ADC sampling rate sets an upper limit on the data transmission rate. Since the ADC in the *TM4C123G* board has a maximum sampling rate of $10^6$ samples/s and to ensure correct decoding of the received signal the data is sampled twice for each bit period setting the maximum possible data rate to 500 Kbps. In addition, the high pass filter cut off frequency of the DC coupling circuit determines the lower threshold of the data rate. If the frequency of operation is too low, due to high pass filtering effect the coupled AC voltage will decay to the dc biasing value, which will result in erroneous decoding of the received signal. The cutoff frequency of the coupling circuit is set at 1 KHz in our implementation. Custom designed comparator could increase the data rate up to tens of Mbps or more.

## IV. MEASUREMENT RESULTS

### A. Signal Transmission Characteristics

The signal transmission characteristics of the intra-body and HMI is primarily determined by the capacitive return path. Therefore, in both cases, the signal transfer characteristics are very similar and we show the waveforms corresponding to the sensor to hub communication. We sample the output from the pulse sensor through an ADC and transmit the received bits through the human body. We observe the signal at 4 points on the system: a) output of the transmitter ($V_{TX}$), b) received signal at the receiver end ($V_{RX}$), c) Integrated received signal ($V_{RX,INT}$), and d) decoded output of receiver.

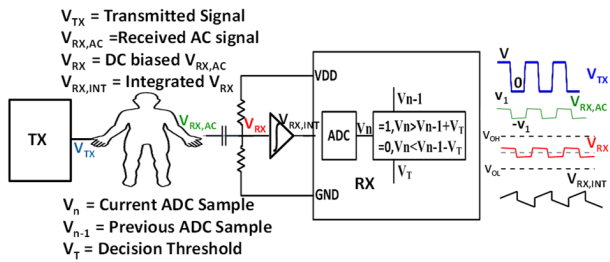

Figure 4: Important signal waveforms for HBC based WHMS

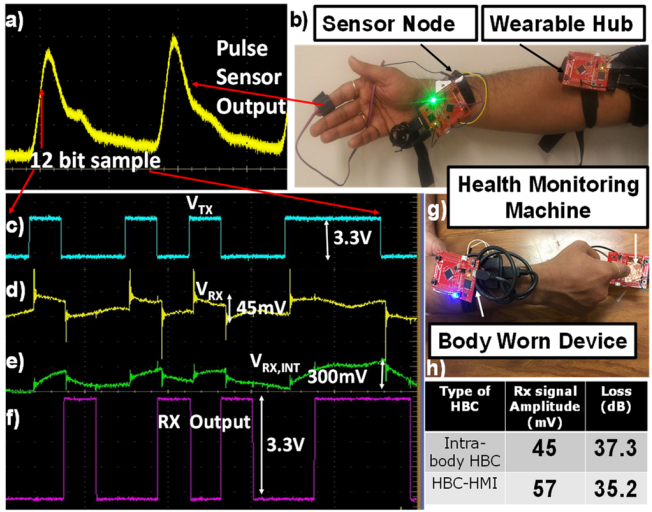

Figure 6: Signal transfer characteristics of HBC based WHMS system. a) Output analog waveform from pulse sensor. b) Diagram showing the communication between a sensor node and WH. c) Transmitted Signal, d) Received Signal with DC biasing, e) Integrated received signal, which is applied to ADC, f) Receiver decoded output. The received signal is ~45 mV which is significantly attenuated from the transmitted value of 3.3V. g) Diagram showing HBC-HMI, where data transmission between a body worn device and a health monitoring machine is done by touching an electrode on the machine. h) Loss and received signal amplitude for a 3.3V transmitted signal in intra-body HBC and HBC-HMI. HBC-HMI shows less loss compared to intra-body HBC.

Figure 6 shows the measured signal characteristics between a health-monitoring node and the hub. The transmitted signal has 3.3 $V_{p-p}$ amplitude (Figure 6c) and is coupled through a copper electrode band into the body and at the receiver end an ac signal with ~45 mV peak to peak amplitude is received (Figure 6d), representing a loss of ~37 dB. Since the transmitter and receiver do not share a common ground, the received signal does not have a dc component in it. The received signal is then coupled onto a DC level of 1.1 V through a coupling circuit, and applied to an integrator, whose output is sampled synchronously through an ADC with a resolution of 0.8 mV. The difference in consecutive sample values are used to determine the logic of the received signal, with logic 1 and logic 0 as 3.3 V and 0V respectively (Figure 6f). The coupling circuit introduces a high pass filter effect, which results in a gradual decay of the received signal towards the DC level. The delay between the decoded received waveform and the transmitted waveform is due to the write delay of the GPIO pins of the microcontroller. The signal loss in **HBC-HMI** (Figure 6g) is lesser compared to intra-body HBC (Figure 6h), but the qualitative signal characteristics remain the same.

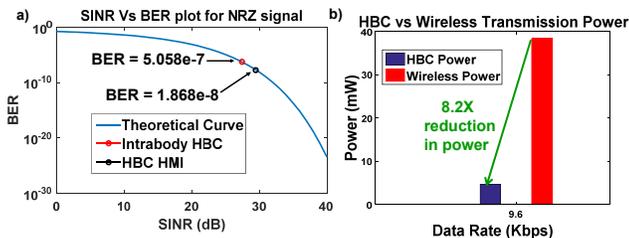

Figure 5: a) Theoretical BER vs SINR plot for NRZ signal, At 130kbps transfer rate there is a SINR of 28 dB and 30dB for intra-body HBC and HBC HMI respectively and a corresponding BER of 5.058e-7 and 1.868e-8. b) Power consumption of HBC based Human-Internet showing almost 8.2X lesser power consumption compared to wireless transmission.

### B. Reliability, Bit Error Rate (BER) of HBC-WHMS

The Bit Error Rate (BER) of a system depends on the Signal to Interference-Noise Ratio (SINR) of the system. The SINR of the current HBC system is limited by the different interference signals present in the environment. The SINR of the signal after integration is 28dB and 30dB for intra-body HBC and HBC-HMI respectively. The theory of efficient interference robust HBC was recently presented in [10][11]. The BER for both intra-body HBC and HBC-HMI are $< 10^{-6}$, as shown in Figure 5a. Hence, the low-loss coupled with the well-controlled wire-like channel characteristics of HBC allows for a very reliable technique for health monitoring from energy-sparse wearable/implantable physiological sensors.

### C. Energy-efficiency of HBC-WHMS

As an indicator of energy efficiency, we compare the power consumption through HBC and Bluetooth data transmission for same data rate. A HC-06 Bluetooth module is used which can transmit only at a baud rate of 9.6 kbps. By sending data at the same rate through HBC, we can see ~8.2X power advantage as shown in Figure 5b. Although the HBC system can support higher data rate, due to the limited data rate support for the Bluetooth module we can compare the power consumption at a single data rate. The energy per bit measurement is limited, due to overhead power consumption in a microcontroller. However, the power efficiency shows the potential benefit of HBC over wireless to a first order. However, a custom ASIC can provide ~100X better energy efficiency with HBC compared to wireless transmission.

## V. CONCLUSION

In this paper, we propose a Capacitive VM HBC based wearable health monitoring system, which enables low energy, secure communication necessary for such energy-constrained systems. A COTS based implementation of such a system is represented and the key design challenges highlighted. The low BER and 8.2X power efficiency of the HBC based system shows its advantage over wireless system.

## VI. ACKNOWLEDGEMENT

This research was supported in part by the National Science Foundation CRII Award (CNS 1657455) and Purdue University Startup Funds.